\documentclass[preprint,aps,floatfix]{revtex4-1}
\usepackage{hyperref}
\everymath={\displaystyle}
\usepackage{young}
\usepackage{rotating}
\usepackage{longtable}
\def\1{\mbox{l\hspace{-0.53em}1}}

\usepackage{multirow}
\newlength{\AccoHaut}

\begin{document}
\title{Negative parity baryons in the $1/N_c$ expansion: the three towers of states revisited}

\author{N. Matagne\footnote{E-mail address: nicolas.matagne@umons.ac.be}}
\affiliation{University of Mons, Service de Physique Nucl\'eaire et 
Subnucl\'eaire,Place du Parc 20, B-7000 Mons, Belgium}

\author{Fl. Stancu\footnote{E-mail address: fstancu@ulg.ac.be}}
\affiliation{University of Li\`ege, Institute of Physics B5, Sart Tilman,
B-4000 Li\`ege 1, Belgium}

\date{\today}

\begin{abstract}
We discuss the compatibility between the quark-shell picture and the meson-nucleon scattering picture
in large $N_c$ QCD for mixed symmetric $\ell$ = 1 states previously analyzed by using a simple Hamiltonian 
including operators up to order $\mathcal{O}(N^0_c)$ defined in the standard ground state symmetric core + excited quark 
procedure. Here we introduce a  Hamiltonian of order $\mathcal{O}(N^0_c)$ defined in a new approach where 
the separation of the system into two parts is not required.  Three degenerate sets of states (towers) with the same
quantum numbers as in the scattering picture and in the standard procedure are obtained.  Thus the 0 is  equally achieved. 
The eigenvalues of the presently chosen Hamiltonian also
have simple analytic expressions, depending linearly on the three dynamical coefficients 
entering the Hamiltonian. 
This reinforces the validity of the new approach
which had  already 0 described excited negative parity baryons in a large energy range.
\end{abstract}

\maketitle

 

\section{Introduction}

Many of the qualitative and quantitative predictions of the
$1/N_c$ expansion method  \cite{HOOFT,WITTEN,Gervais:1983wq,DM},  where $N_c$ is the number of colors,  
have been proved 0 when compared to experimental results 
of ground state baryons \cite{Jenk1,DJM94,DJM95,CGO94}.
The method is based on the observation that, for $N_f$ flavors, the ground state baryons display an exact 
contracted SU(2$N_f$) symmetry when $N_c \rightarrow \infty$.  At large, but finite $N_c$, 
this symmetry is broken by contributions of order of $1/N_c$,  leading to mass splittings.  
 
Subsequently, efforts have been made to extend the method to excited states. There are two complementary 
pictures of large $N_c$ for baryon resonances. One is the so-called {\it quark-shell picture} where the 
symmetry is extended to SU(2$N_f$) $\times$ O(3), which allows  to
classify baryons in excitation bands $N$  \cite{Matagne:2005gd}, like in the quark model where $N_c$ = 3 \cite{Semay:2007cv}. 
The other is the {\it resonance or scattering picture} 
derived from symmetry features of chiral soliton models. 
The role of large $N_c$ QCD is to relate the scattering amplitudes in various channels with $K$-amplitudes, where $K$
is the grand spin $\vec{K} = \vec{I} + \vec{J}$. 
These are  linear relations in the meson-nucleon scattering amplitudes from which one can infer some patterns of 
degeneracy among resonances. 

Naturally, there has been concern about the compatibility of the two pictures. 
Simultaneously Cohen and Lebed   \cite{COLEB1} and  Pirjol and Schat \cite{Pirjol:2003ye} studied the $N$ = 1 
band which represents  the lowest $[70,1^-]$ multiplet and found 
the same three sets of degenerate states as in the {\it resonance or scattering picture}. 
In both studies there were three leading order operators in the mass formula, 
$c_1 \ \1$ of order  $\mathcal{O}(N_c)$ together with  $\ell \cdot s$  and 
$\frac{1}{N_c}\ell^{(2)} \cdot g \cdot G_c$ having matrix elements of order $\mathcal{O}(N^0_c)$.
In particular, Pirjol and Schat showed that the three sets of degenerate states 
correspond to irreducible representations of the contracted SU(4)$_c$
symmetry, the three sets  being called  three towers of states.
Moreover, to the three leading order operators in the mass formula they added $1/N_c$
corrections and reanalyzed the mass spectrum of the lowest negative parity nonstrange 
baryons.  They found  ambiguities in the identification of physical states with $N_c$ = 3 
with the degenerate large $N_c$ tower states. Actually, in the SU(4) case the degeneracy of 
sets of states corresponding 
to irreducible representations of the contracted SU(4)$_c$ symmetry
was first discussed  by Pirjol and Yan in Ref. \cite{PY2}.

Later on, the compatibility between the two pictures was discussed  on a general basis   
again by Cohen and Lebed  \cite{COLEB2}. 
The compatibility was  claimed for completely symmetric (S), mixed symmetric (MS) and completely
antisymmetric (A) states of $N_c$ quarks having angular momentum up to $\ell = 3$.  
In Ref.  \ \cite{Matagne:2011sn} we gave 
an explicit proof of the degeneracy of mass eigenvalues in the quark-shell picture  for $\ell = 3$. 
We thus supported  the idea of full compatibility of Ref. \cite{COLEB2} for higher parts of the spectrum.
This compatibility means that any complete 
spin-flavor multiplet within one picture fills the quantum numbers of the
other picture. In addition, we could prove that the quark-shell picture is richer in information, by making a clear distinction 
between degenerate sets of states of different values of the angular momentum but
associated to the same grand spin $K$. For example, one can associate a common $K$ = 2 
to both $\ell$ = 1 and $\ell$ = 3.  A similar situation appears for every $K$ value associated 
to two distinct values of $\ell$ satisfying the $\delta(K \ell 1)$ rule \cite{COLEB2}.

It is important to mention that the above studies were based  on the procedure  
to consider an excited state as a single quark excitation about a ground state 
symmetric core \cite{Goi97,CCGL}.  In fact the operators mentioned above follow the notation 
of Refs. \cite{Goi97,CCGL}, namely lower case indicates operators acting on the excited quark 
and the subscript $c$  is attached to those   acting on the symmetric core. 

The symmetric core + excited quark was originally proposed \cite{Goi97} as an extension of the 
ground state treatment to excited states and was inspired by the Hartree picture. 
In this way, in the flavor-spin space,  the problem was 
reduced to the knowledge of matrix elements of the SU(2N$_f$) generators between symmetric 
states, already known from the ground state studies. Accordingly, the wave function was approximately 
given by the coupling of an excited quark to a ground state core of $N_c-1$ quarks, 
without performing antisymmetrisation.  0 
the symmetric core + an excited quark approach was  supported by Pirjol and Schat \cite{Pirjol:2007ed}
within a general large $N_c$ constituent quark model  Hamiltonian
starting from an exact wave function.
A practical problem is that the number of operators entering the mass formula is exceedingly large. 
For example, for the $[70,1^-]$ multiplet in SU(4) there are 12 linearly independent operators in 
powers of $1/N_c$ included, while in the $N$ = 1 band there are seven experimentally known resonances
+ two mixing angles. It is therefore difficult, if not impossible, to find out the most dominant operators.
In addition the symmetric core always has equal spin and isospin, therefore some information is lost
regarding the baryons structure. 

As an alternative, in Ref.  \cite{Matagne:2006dj}  we have proposed a method where 
all identical quarks are treated on the same footing and we have an exact wave 
function in the orbital-flavor-spin space. The procedure has been 0 applied to the 
$N$ = 1 band  \cite{Matagne:2006dj,Matagne:2008kb,Matagne:2011fr}   and recently to the $N$ = 3 band 
\cite{Matagnenew},  where data are very scarce. 
We found out that the most dominant operators of order $1/N_c$ were both the spin and flavor, the latter being neglected in all studies 
based on the symmetric core + an excited quark approach.  They  are $\frac{1}{N_c}~S^i S^i$ and 
$\frac{1}{N_c}[T^a T^a - \frac{1}{12}N_c(N_c+6)]$ respectively , where the latter is compatible 
both with SU(4) and SU(6) \cite{Matagne:2008kb}. The generators $S^i$ and $T^a$ act on the whole system.

It is precisely in our approach that we wish to analyze the compatibility of the {\it quark-shell picture} and of the
{\it resonance or scattering picture}. It will be shown that
the present results are as simple as those of the symmetric core + an excited quark approach.
It is remarkable that in the {\it quark-shell picture} 
described below the Hamiltonian eigenvalues are  linear analytic functions of the dynamical 
coefficients $c_i$ entering the Hamiltonian, like in Refs.  \cite{COLEB1,Pirjol:2003ye}. 
In this way one can easily identify the
three sets of degenerate states which are identical in the quantum numbers with those
of the {\it resonance or scattering picture} and those of Refs. \cite{COLEB1,Pirjol:2003ye}.
The compatibility is therefore confirmed and this gives strong support to our procedure. 

\section{A simplified mass operator }

In the  quark-shell  picture  the leading-order Hamiltonian corresponding to the procedure of 
Refs.  \cite{Matagne:2006dj,Matagne:2008kb,Matagne:2011fr}   
has the following form 
\begin{equation}\label{TOY}
H = \sum c_i O_i,
\end{equation}
where the operators presently under consideration are 
\begin{equation}
O_1 = N_c \ \1 , ~~~  O_2 = \ell \cdot s, ~~~   O_6 =  \frac{15}{N_c} L^{(2)ij} G^{ia} G^{ja},
\end{equation}
in the notations of Ref. \cite{Matagne:2011fr} for the operators $O_i$.
The first two terms are identical to those of Refs.  \cite{COLEB1} or \cite{Pirjol:2003ye}. 
The matrix elements of the first term are $N_c$ on all baryons and the second term 
is a one-body operator defined in the spirit of the Hartree picture \cite{WITTEN}
and its matrix elements are of order $\mathcal{O}(N^0_c)$.  
The  third term  is  new and consistent with our procedure described in the introduction.
It is a two-body operator but has  matrix elements of order $\mathcal{O}(N^0_c)$. 
It contains the tensor  
$L^{(2)ij}$ of SO(3) defined as
\begin{equation}\label{TENSOR} 
L^{(2)ij} = \frac{1}{2}\left\{L^i,L^j\right\}-\frac{1}{3}
\delta_{i,-j}\vec{L}\cdot\vec{L},
\end{equation}
which acts on the orbital wave function  
of the whole system of $N_c$ quarks and is normalized as in Ref. \cite{Matagne:2005gd}. Note that 
when the angular momentum acts on the whole system we use capital $L_i$
to distinguish it from $\ell_i$, in the spin-orbit operator, which acts on a single quark. 
In our approach $O_2$ and $O_6$ are the only operators of order $\mathcal{O}(N^0_c)$.
The neglect of $1/N_c$
corrections  in the $1/N_c$ expansion makes sense for the comparison with the scattering picture
in the large $N_c$ limit, as already discussed in Ref. \cite{COLEB1}.

\section{The basis states}

The Hamiltonian (\ref{TOY}) is diagonalized in the bases described below.

\subsection{The nucleon case}
We have the following $[N_c-1,1]$ spin-flavor  ($SF$) states which form a symmetric state with the orbital $\ell$ = 1 state of partition $[N_c - 1,1]$ as well  

\begin{equation}\label{n1} 
\left[N_c - 1, 1\right]_{SF} = \left[\frac{N_c+1}{2}, \frac{N_c - 1}{2}\right]_{S} \times  
\left[\frac{N_c+1}{2}, \frac{N_c - 1}{2}\right]_{F}, 
\end{equation}
where $N_c  \geq 3$ and $S = 1/2$, $J = 3/2$,
\begin{equation}\label{n2}
\left[N_c - 1, 1\right]_{SF} = \left[\frac{N_c+3}{2}, \frac{N_c - 3}{2}\right]_{S} \times  
\left[\frac{N_c+1}{2}, \frac{N_c - 1}{2}\right]_{F}, 
\end{equation} 
where $N_c  \geq 3$   and $S = 3/2$, $J = 1/2, 3/2, 5/2$. As one can easily see,
they give rise to matrices associated to a given $J$ which are either $2 \times 2$ or $1 \times 1$.

\subsection{The $\Delta$ case}
We have the following basis states in the spin-flavor space compatible with the orbital state $[N_c -1,1]$ with $\ell$ = 1

\begin{equation}\label{d1}
\left[N_c - 1, 1\right]_{SF} = \left[\frac{N_c+1}{2}, \frac{N_c - 1}{2}\right]_{S} \times  
\left[\frac{N_c+3}{2}, \frac{N_c - 3}{2}\right]_{F} ,  
\end {equation}
where  $N_c  \geq 3$ and $S = 1/2$, $J = 3/2$, denoted in the following as  $^210_J$, 
\begin{equation}\label{d2}
\left[N_c - 1, 1\right]_{SF} = \left[\frac{N_c+3}{2}, \frac{N_c - 3}{2}\right]_{S} \times \left[\frac{N_c+3}{2}, \frac{N_c - 3}{2}\right]_{F} ,   
\end {equation}
 where   $N_c  \geq 5$ and   $S = 3/2$,  $J = 1/2, 3/2, 5/2$,  denoted in the following as  $^410_J$,
\begin{equation} \label{d3}
\left[N_c - 1, 1\right]_{SF} = \left[\frac{N_c+5}{2}, \frac{N_c - 5}{2}\right]_{S} \times 
\left[\frac{N_c+3}{2}, \frac{N_c - 3}{2}\right]_{F},   
\end{equation}
 where $N_c  \geq 7$  and $S = 5/2$ and $J = 1/2, 3/2, 5/2, 7/2$,  denoted in the following as  $^610_J$ .

For $N_c = 3$ the first state belongs to the $^210$ multiplet.
The other two types of states do not appear in the real world with $N_c = 3$.
As above, it is easy to find out the size of the matrix of a fixed $J$. For example for $J = 3/2$ we have a 3 $\times$ 3 matrix 
defined in the space of all $\Delta$ states.

\section{Results and discussion}

\begin{table*}[h!]
\begin{center}
\caption{Diagonal matrix elements of $O_1$, $O_2$ and $O_6$ for all states belonging to the 
$[{\bf 70},1^-]$ multiplet.  }
\label{Matrix}
\renewcommand{\arraystretch}{2.3} {\scriptsize
\begin{tabular}{lcccccccccc}
\hline
\hline
  &\hspace{0cm}  $O_1$ &\hspace{0cm} $O_2$ 
  & \hspace{0cm}  $O_6$  
  &\hspace{0cm}     \\
  \hline
$^28_{\frac{1}{2}}$ & $N_c$  & $-\frac{2N_c-3}{3N_c}$  
&   
    $0$ &\\
$^48_{\frac{1}{2}}$ & $N_c$   &  $-\frac{5}{6}$ 
  & $-\frac{25(N_c-1)}{8N_c}$ & \\
$^28_{\frac{3}{2}}$ & $N_c$  & $\frac{2N_c-3}{6N_c}$ 
 &  $0$  &\\
$^48_{\frac{3}{2}}$ & $N_c$  &  $-\frac{1}{3}$  
  & $\frac{5(N_c-1)}{2N_c}$   &\\
$^48_{\frac{5}{2}}$ & $N_c$ & $\frac{1}{2}$  
  & $-\frac{5(N_c-1)}{8N_c}$ &\\
$^2{10}_{\frac{1}{2}}$ & $N_c$ & $\frac{1}{3}$ 
  & $0$ & \\
$^2{10}_{\frac{3}{2}}$ & $N_c$ & $-\frac{1}{6}$  
  & $0$ &\\
$^4{10}_{\frac{1}{2}}$ & $N_c$ & $-\frac{1}{6}$ 
  & $ \frac{5(N_c+5)}{2N_c}$ &\\
$^4{10}_{\frac{3}{2}}$ & $N_c$ & $-\frac{2}{15}$  
  & $ -\frac{2(N_c+5)}{N_c}$ &\\    
$^4{10}_{\frac{5}{2}}$ & $N_c$ & $\frac{1}{5}$  
  & $ \frac{N_c+5}{2N_c}$ &\\    
$^6{10}_{\frac{3}{2}}$ & $N_c$ & $-\frac{7}{10}$  
  & $-\frac{7(3 N_c-25)}{12 N_c}$ &\\
$^6{10}_{\frac{5}{2}}$ & $N_c$ & $-\frac{1}{5}$  
  & $\frac{2(3 N_c-25)}{3 N_c}$ &\\
$^6{10}_{\frac{7}{2}}$ & $N_c$ & $\frac{1}{2}$  
  & $-\frac{5(3 N_c-25)}{24 N_c}$ &\\
\hline \hline
\end{tabular}}
\end{center}
\end{table*}

\begin{table*}[h!]
\begin{center}
\caption{Off-diagonal matrix elements of $O_1$, $O_2$ and $O_6$ for all states belonging to the 
$[{\bf 70},1^-]$ multiplet.  }
\label{Off-diag}
\renewcommand{\arraystretch}{2.3} {\scriptsize
\begin{tabular}{lcccccccccc}
\hline
\hline
  &\hspace{0cm}  $O_1$ &\hspace{0cm} $O_2$ 
  & \hspace{0cm}  $O_6$  
  &\hspace{0cm}     \\
  \hline  
$^28_{\frac{1}{2}} -$ $^48_{\frac{1}{2}}$ & 0 & $-\frac{1}{3}\sqrt{\frac{N_c+3}{2N_c}}$ &
$-\frac{25}{4}\sqrt{\frac{N_c+3}{2 N_c}}$ & \\
$^28_{\frac{3}{2}}-$ $^48_{\frac{3}{2}}$ & 0 & $-\frac{1}{6}\sqrt{\frac{5(N_c+3)}{N_c}}$ & 
$\frac{5}{8}\sqrt{\frac{5(N_c+3)}{N_c}}$ & \\
$^210_{\frac{1}{2}}-$ $^410_{\frac{1}{2}}$ & 0 & $\frac{1}{6}\sqrt{\frac{5(N_c-3)}{N_c}}$ & 
$-\frac{5}{8}\sqrt{\frac{5(N_c-3)}{N_c}}$ & \\
$^210_{\frac{3}{2}}-$ $^410_{\frac{3}{2}}$ & 0 & $\frac{5}{6}\sqrt{\frac{N_c-3}{2N_c}}$ & 
$\frac{5}{8}\sqrt{\frac{(N_c-3)}{2N_c}}$ & \\
$^210_{\frac{3}{2}}-$ $^610_{\frac{3}{2}}$ & 0 & $ 0$ & 
$\frac{5}{8N_c}\sqrt{3(N_c-3)(N_c+5)}$ & \\
$^410_{\frac{3}{2}}-$ $^610_{\frac{3}{2}}$ & 0 & $ -\frac{3\sqrt{6}}{20}$ & 
$-\frac{21}{16}\sqrt{\frac{6(N_c+5)}{N_c}}$ & \\
$^410_{\frac{5}{2}}-$ $^610_{\frac{5}{2}}$ & 0 & $ -\frac{\sqrt{21}}{10}$ & 
$\frac{3}{8}\sqrt{\frac{21(N_c+5)}{N_c}}$ & \\
\hline \hline
\end{tabular}}
\end{center}
\end{table*}

The analytic expressions of the resulting matrix elements of the operators contained in the Hamiltonian (\ref{TOY}) 
are given in Tables  \ref{Matrix} and \ref{Off-diag},  as a function of $N_c$.   
Details of the calculation of the matrix elements  are given in  Appendix  A.

To obtain the matrices to be diagonalized we have to take the limit $N_c \rightarrow \infty$ 
in the matrix elements of $O_2$ and $O_6$ given in  Tables \ref{Matrix} and \ref{Off-diag}.
As an example we present the largest possible matrix,  corresponding to  $\Delta_{3/2}$ states, which is
\begin{eqnarray}
\label{delta5halves}
M^{\ell=1}_{\Delta_{3/2}} & = &
 \left( 
 \begin{array}{ccc}
c_1 N_c - \frac{1}{6} c_2  &~~   \frac{5}{\sqrt{2}} \left( \frac{c_2}{6}  +  \frac{c_6}{8} \right) 
&~~ \frac{15 \sqrt{3} }{8}  c_6 \\
 \frac{5}{\sqrt{2}} \left( \frac{c_2}{6}  +  \frac{c_6}{8}\right)    &~~  c_1 N_c - \frac{2}{15} c_2 -  2 c_6 
  &~~   - \frac{3 \sqrt{6}}{20} c_2 -   \frac{21 \sqrt{6}}{16}c_6 \\ 
 \frac{15 \sqrt{3}}{8} c_6     &~~    - \frac{3 \sqrt{6}}{20} c_2 -   \frac{21 \sqrt{6}}{16 }c_6 
  &~~   c_1 N_c - \frac{7}{10} c_2  - \frac{7}{4} c_6
\end{array} 
\right),
\end{eqnarray}
The eigenvalues of this matrix are 
\begin{equation}\label{m0} 
m'_0 = c_1 N_c - c_2 - \frac{25}{4} c_6,
\end{equation}
\begin{equation}\label{m1} 
m'_1 = c_1 N_c - \frac{1}{2} c_2 + \frac{25}{8} c_6,
\end{equation}
\begin{equation}\label{m2} 
m'_2 = c_1 N_c + \frac{1}{2} c_2 - \frac{5}{8} c_6.
\end{equation}
The other matrices follow straightforwardly from Tables  \ref{Matrix} and \ref{Off-diag}. By diagonalizing all matrices
we found that the eigenvalues (\ref{m0})-(\ref{m2})
are the only possible ones  for all $N_J$ and $\Delta_J$ presented above.
Accordingly  the following sets of degenerate negative parity multiplets were found for $\ell$ = 1 orbital 
excitations 
\begin{equation}\label{K0}
N_{1/2}, ~ \Delta_{3/2}, 
 ~~~~~(m'_0)
\end{equation} 
\begin{equation}\label{K1}
N_{1/2},~ \Delta_{1/2},~N_{3/2},~ \Delta_{3/2},~ \Delta_{5/2}, 
 ~~~~~(m'_1)
\end{equation} 
\begin{equation}\label{K2}
\Delta_{1/2},  ~ N_{3/2},   ~ \Delta_{3/2}, 
~ N_{5/2},  ~ \Delta_{5/2},   
~ \Delta_{7/2}, 
 ~~~~~(m'_2)
\end{equation} 
where,  on the right side we indicate the mass of each degenerate set.
These degenerate multiplets are identical to those found in Refs. \cite{COLEB1} and 
\cite{Pirjol:2003ye}. The masses $m'_i$ of Eqs. (\ref{m0})-(\ref{m2}) are naturally different from $m_i$ of 
the above references because the Hamiltonian is different in structure, it contains different 
dynamical coefficients but has
similar large $N_c$ properties.  Simple forms as those of Eqs. (\ref{m0})-(\ref{m2})
hold only for a Hamiltonian of order $\mathcal{O}(N^0_c)$. Other choices within 
the procedure of  Refs.  \cite{Matagne:2006dj,Matagne:2008kb,Matagne:2011fr}   
would lead to the inclusion of $1/N_c$ corrections which will necessarily imply numerical 
calculations. 

Another remarkable aspect of the present study concerns the mixing angles. 
For those sectors for which large $N_c$ and $N_c$ = 3 have the same number of quark 
model states, which, in particular  is  the nucleon case with $J$ = 1/2 and $J$ = 3/2,
see Eqs.  (\ref{n1}) and (\ref{n2}), the mixing angles are identical to 
those obtained by Cohen and Lebed \cite{COLEB1} or Pirjol and Schat \cite{Pirjol:2003ye}.
This means $\tan \theta_{N_{1/2}} = \sqrt{2}$ and
 $\tan \theta_{N_{3/2}}  = - \frac{1}{\sqrt{5}}$ and amounts to 
 $\theta_{N_{1/2}}$ = 0.96 rad and  $\theta_{N_{3/2}}$ = 2.72 rad respectively (for details see 0 B).
The mixing angles determined from fits to $N^*$ strong decays and 1 
data are $\theta_{N_{1/2}}  =  0.39 \pm 0.11$ rad and 
$\theta_{N_{3/2}} = 2.82 \pm 0.11$ rad, respectively \cite{Goity:2004ss,Scoccola:2007sn}.
This suggests that for $N_{3/2}$ states the agreement with the phenomenological
value is nearly achieved at order $\mathcal{O}(N^0_c)$ but for  $N_{1/2}$ states
corrections of order $\mathcal{O}(N^{-1}_c)$ are necessary. 
In Ref. \cite{COLEB1} the comparison has been made with the decay data of Ref. \cite{HEY},
from where it has been extracted $\theta_{N_{1/2}}$ = 0.56 rad.

For the other mixing angles resulting from $2 \times 2$ matrices we find 
$\tan \theta_{\Delta_{1/2}}  =  \sqrt{1/5}$ and 
$\tan \theta_{\Delta_{5/2}}  = - \sqrt{3/7}$. 
These give the same absolute values for the mixing angle as those of Cohen and Lebed \cite{COLEB1}
but of opposite signs. It may be a matter of phase convention.

We also have to compare these results with those of the meson-nucleon scattering picture,
where linear relations between matrix elements  $S^{\pi}_{LL'RR'IJ}$ and  $S^{\eta}_{LRJ}$ of $\pi$ and $\eta$ scattering off 
a ground state baryon in terms of $K$-amplitudes were derived. They are given by the following equations 
\begin{equation}\label{pi}
S^{\pi}_{LL'RR'IJ} = \sum_K ( - 1)^{R'-R} \sqrt{(2R+1)(2R'+1)} (2K+1)
\left\{\begin{array}{ccc}
        K& I & J \\
	R' & L' & 1
      \end{array}\right\} 
 \left\{\begin{array}{ccc}
        K& I & J \\
	R & L & 1
      \end{array}\right\}  
      s^{\pi}_{KLL'}    
\end{equation}
and
\begin{equation}\label{eta}
S^{\eta}_{LRJ} = \sum_K \delta_{KL}\delta(LRJ) s^{\eta}_{K},
\end{equation}
in terms of the reduced amplitudes $s^{\pi}_{KL'L}$ and $s^{\eta}_{K}$ respectively. 
These equations were first derived in the context  of the chiral soliton model 
\cite{HAYASHI,MAPE,MATTIS,MattisMukerjee}
where 
the mean-field breaks the rotational and isospin symmetries, so that $J$ and $I$ are not
conserved but the ${\it grand}$ ${\it spin}$  $K$ is conserved and excitations can be labelled by $K$.
These relations are exact in large $N_c$ QCD and are independent of any model assumption.

The explicit form of these equations can be found in Table I of Ref. \cite{COLEB1}. That table infers  
a pattern of degeneracy identical to that presented in  Eqs. (\ref{K0})-(\ref{K2}).
The contributing amplitudes are $s^{\eta}_0$ for the resonances listed in Eq.  (\ref{K0}),
$(s^{\pi}_{100}, s^{\pi}_{122})$ for those of  Eq. (\ref{K1}) and  $(s^{\pi}_{222}, s^{\eta}_{2})$ for those of Eq. (\ref{K2})
(for details see Ref. \cite{COLEB1}). In the resonance picture the degenerate towers of states 
(\ref{K0})-(\ref{K2})
correspond to the grand spin $K$ = 0,1 and 2 respectively.  Therefore the triangular rule proposed
in Ref. \cite{COLEB2} is satisfied. 


\section{Conclusions}
The compatibility between the quark-shell picture of the $1/N_c$ expansion method
and the meson-nucleon resonance picture 
has been previously analyzed  \cite{COLEB1,Pirjol:2003ye,COLEB2} by starting from a Hamiltonian 
containing operators of order  $\mathcal{O}(N^0_c)$ 
defined in the symmetric core + an excited quark method \cite{Goi97,CCGL} and full compatibility 
has been found. 
Here we have used an alternative description of the mixed symmetric states where the 
separation of SU(6) generators into two terms, one  acting on the core and the other on the excited quark, 
is avoided  and the orbital-flavor-spin  wave function is exactly symmetric under the permutation group 
\cite{Matagne:2006dj,Matagne:2011fr}, as it should be for identical quarks.
Interestingly we found an identical pattern of degeneracy in the quantum numbers with that obtained from the
symmetric core + an excited quark method, thus have proven that the full compatibility
holds in this procedure as well.  This supports once more  the method we have proposed 
in Refs.  \cite{Matagne:2006dj,Matagne:2011fr} where a good fit to the experiment 
has been found for the $N$=1 band, and more recently for mixed symmetric multiplets in the
$N$=3 band \cite{Matagnenew}.

The importance of the compatibility of the two pictures has been clearly pointed out in Ref. \cite{COLEB1}
where it was also stressed that it does not justify all aspects of the 
{\it quark-shell picture}, in particular the dynamical details, but it justifies those aspects
of the model that essentially follow from the contracted SU(2$N_f$) symmetry.

\appendix

\section{}       

For the calculation of the matrix elements of the spin-orbit operator $O_2 =  \ell \cdot s$ one should refer
to the Appendix of Ref. \cite{Matagne:2011sn} where the notations of Ref. \cite{Stancu:1991rc}
were used for the isoscalar factors of the permutation group.  In Ref. \cite{Matagne:2011sn}
the matrix elements of $O_2$ for 
orbital excitation with $\ell$ = 3 have been calculated.

For the   reader's benefit here we give some useful details for the calculation of the 
matrix elements of the operator $O_6 = 15/N_c~ L^{2} \cdot G \cdot G$. The expression of the matrix elements of    
the operator   $L^{2} \cdot G \cdot G$   as obtained  in Ref.        \cite{Matagne:2011fr}    is                     
 \begin{eqnarray}
 \lefteqn{  \langle (\lambda'\mu')Y'I'I_3';\ell'S' JJ_3
 |(-1)^{i+j+a}L^{(2)ij}G^{-ia}G^{-j,-a}|(\lambda\mu)YII_3;\ell SJJ_3\rangle =  }\\
 & &    \delta_{\ell'\ell}\delta_{\lambda\lambda'}\delta_{\mu\mu'}
 \delta_{Y'Y} \delta_{I'I} \delta_{I_3'I_3} (-1)^{J+\ell -S}
  \frac{1}{2}C^{SU(6)}_{[f]} \sqrt{\frac{5\ell(\ell+1)(2\ell-1)(2\ell+1)(2\ell+3)}{6}} \nonumber \\
 & \times  & \sqrt{(2S+1)(2S'+1)}
  \left\{\begin{array}{ccc}
   \ell & \ell & 2 \\
      S & S' & J
  \end{array}\right\}
     \sum_{S''}(-1)^{(S-S'')}\left\{\begin{array}{ccc}
   1 & 1 & 2 \\
   S & S' & S''
  \end{array}\right\} 
  \nonumber \\ &\times  & \sum_{\rho,\lambda'',\mu''}
   \left(\begin{array}{cc||c}
         [f] & [21^4] & [f] \\
	 (\lambda''\mu'')S'' & (11)1 & (\lambda\mu)S 
        \end{array}\right)_{\rho} 
    \left(\begin{array}{cc||c}
         [f] & [21^4] & [f] \\
	 (\lambda''\mu'')S'' & (11)1 & (\lambda\mu)S' 
        \end{array}\right)_{\rho},
 \end{eqnarray} 
where the partition of mixed symmetric states under consideration is $[f] = [N_c-1,1]$ and 
\begin{equation}
C^{SU(6)}_{[f]} = \frac{N_c(5N_c+18)}{12},
\end{equation}
is the Casimir operator of SU(6) for the  partition $[f]$.
The  two factors appearing in the sum over $\rho$, $\lambda''$ and $\mu''$
  are isoscalar factors of SU(6) to be found in Ref.  \cite{Matagne:2011fr} as well.
As an example we consider the case of $\Delta_{3/2}$ states defined by Eqs. (\ref{d1})-(\ref{d3}).
For states of type (\ref{d1}),  (\ref{d2})  and (\ref{d3}) the isoscalar factors should be taken from Table VII,  
V and  VI respectively.  These tables are very general and 
the meaning is as follows. Table VII corresponds to spin-flavor (FS) states where the Young diagram associated to spin
has $\lambda-2$ boxes  in the first row  if the Young diagram associated to flavor has $\lambda$ boxes, as shown below for $N_c = 7$, $(\lambda, \mu) = (3,2)$  
\begin{eqnarray}
\label{}
\raisebox{-9.0pt}{\mbox{\begin{Young}
 & & & & & \cr
 \cr
\end{Young}}}\ \ & = & 
\raisebox{-9.0pt}{\mbox{\begin{Young}
& & & \cr
& & \cr
\end{Young}}}\ \ \times
\raisebox{-9.0pt}{\mbox{\begin{Young}
& & & & \cr
&  \cr
\end{Young}}}
\end{eqnarray}

 The multiplet $^210$ from
the real world ($N_c=3$) is a particular case. It has one  box in the first row of the Young diagram corresponding to spin, which gives $S=1/2$ and three 
boxes in the Young diagram corresponding to flavor,  which gives  $(\lambda, \mu) = (3,0)$, describing a $\Delta$ state.

Table V corresponds to FS states where the number of extra boxes in the first row, both in the spin and flavor diagrams is the same,
namely three, as shown below for $N_c = 7$
\begin{eqnarray}
\label{}
\raisebox{-9.0pt}{\mbox{\begin{Young}
 & & & & & \cr
 \cr
\end{Young}}}\ \ & = & 
\raisebox{-9.0pt}{\mbox{\begin{Young}
& & & & \cr
&  \cr
\end{Young}}}\ \ \times
\raisebox{-9.0pt}{\mbox{\begin{Young}
& & & & \cr
&  \cr
\end{Young}}}
\end{eqnarray}
That is  why we denoted these spurious states ($N_c \ge 5)$  by $^410_J$. In the real world,
$N_c=3$, this case is realized for $^28$ states with one extra box in both spin and flavor spaces, which means $S= 1/2$ and $(\lambda,\mu) = (1,1)$.

  Table VI corresponds to 
FS states where the number of extra boxes is two units larger in the spin space than in the flavor space, see below for $N_c = 7$
\begin{eqnarray}
\label{}
\raisebox{-9.0pt}{\mbox{\begin{Young}
 & & & & & \cr
 \cr
\end{Young}}}\ \ & = & 
\raisebox{-9.0pt}{\mbox{\begin{Young}
& & & & & \cr
 \cr
\end{Young}}}\ \ \times
\raisebox{-9.0pt}{\mbox{\begin{Young}
& & & & \cr
&  \cr
\end{Young}}}
\end{eqnarray}
Then for $\Delta$ states one must have $S$ = 5/2 which implies the notation $^610$ for these states.
The real case, $N_c=3$, is the $^48$ multiplet.  

All real cases are indicated in the headings of
 Tables VII, V and VI respectively of Ref. \cite{Matagne:2011fr}.

\section{}

The mixing angles extracted from electromagnetic and strong decays are defined as
\begin{eqnarray}
|N_J(\mathrm{upper}) \rangle = \cos \theta_J |^4N_J \rangle +
 \sin \theta_J |^2N_J \rangle, \nonumber \\
|N_J(\mathrm{lower}) \rangle =  
- \sin \theta_J |^4N_J \rangle +  \cos \theta_J |^2N_J \rangle ,
\end{eqnarray}
where  $|^4N_J \rangle$ and $|^2N_J \rangle$ are the initial states 
in the quark model basis and ${\it upper}$ and ${\it lower}$ are   the physical states 
with upper and lower energies. Here the mixing is due to the spin-orbit 
and to the $O_6$ operator because they both have off-diagonal matrix elements, 
see Table \ref{Off-diag}.   As an example we show the matrix of the $N_{1/2}$ states

\begin{eqnarray}
\label{onehalf}
M^{\ell=1}_{N_{1/2}} & = &
 \left( 
 \begin{array}{cc}
c_1 N_c -\frac{2}{3} ~c_2  &~~~~~~ - \frac{1}{3 \sqrt{2} }~ c_2 -  \frac{25}{4 \sqrt{2}} ~c_6 \\
- \frac{1}{3 \sqrt{2} } ~c_2 -  \frac{25}{4 \sqrt{2}} ~c_6  &~~~~~~   c_1 N_c - \frac{5}{6} ~c_2 + \frac{25}{6} ~c_6  
\end{array} 
\right), 
\end{eqnarray}   
This suggests that the general form of a 2 $\times$ 2 matrix to be diagonalized is
\begin{eqnarray}
\label{N}
M^{\ell}_{N_{J}} & = &
 \left( 
 \begin{array}{cc}
A &~~~~~~ B \\
B  &~~~~~~ C
\end{array} 
\right), 
\end{eqnarray}  
so the mixing angle turns out to be
\begin{equation}      
\tan 2 \theta = - \frac{2B}{C-A}.           
\end{equation}    
Then, for example, from Eq. (\ref{onehalf}) it follows that  $\tan \theta_{N_{1/2}} = \sqrt{2}$. 


\end{document}